# A LOW-DELAY REFERENCE TRACKING ALGORITHM FOR MICROWAVE MEASUREMENT AND CONTROL*

J.F. Zhu[†], H.L. Ding[1][‡], H.K. Li[1], J.W. Han, X.W. Dai, Z.C. Chen[1], J.Y. Yang[1], W.Q. Zhang[1]
Institute of Advanced Science Facilities, Shenzhen, China
[1]also at Dalian Institute of Chemical Physics, Chinese Academy of Sciences, Dalian, China

*Abstract*

In FEL (Free-Electron Laser) accelerators, LLRF (Low-Level Radiofrequency) systems usually deploy feedback or feedforward algorithms requiring precise microwave measurement. The slow drift of the clock allocation network of LLRF significantly impacts the measured microwave phase, thereby affecting the stability of the closed-loop operation. The reference tracking algorithm is used to eliminate the measurement drift. The conventional algorithm is to perform phase and amplitude demodulation on the synchronous reference signal from the main oscillator and subtract the reference phase in other measurement channels. The demodulation is usually based on the CORDIC, which requires approximately 16 clock cycles in FPGA (Field Programmable Gate Arrays). This paper uses the multiplication of complex numbers, which only requires four clock cycles of computational delay and achieves phase subtraction point by point. However, experiments show that it causes irrelevant amplitude noise to overlap and increase the amplitude measurement noise. Nevertheless, this reference tracking algorithm is suitable for control algorithms with low-delay requirements of microwave measurement.

## LOW-DELAY REFERENCE TRACKING ALGORITHM

The reference tracking algorithm eliminates the measurement drift in LLRF (Low-Level Radiofrequency) microwave detection [1, 2]. Figure 1 describes a typical LLRF system consisting of a microwave detector, controller, and actuator. The synchronization system generates a reference signal for LLRF. A drift exists between the synchronization system, the microwave detector, and the actuator. The clock allocation network of LLRF, which consists of a clock local oscillator and RF front-end, mainly causes the drift.

As shown in Fig. 1, both the reference signal and klystron forward are detected, and their amplitude and phase are demodulated. The following formula is the general reference tracking algorithm. We can remove the measurement drift from the reference signal by phase subtraction, as shown in Eq. (1).

$$\varphi'_{mea} = \varphi_{mea} - \varphi_{ref} \quad (1)$$

However, the demodulation is usually based on the CORDIC, which requires approximately 16 clock cycles in FPGA (Field Programmable Gate Arrays) to perform an accuracy of more than 0.002 degrees. This paper uses the multiplication of complex numbers, which only requires four clock cycles of computational delay and achieves phase subtraction point by point. As the equation demonstrates, our algorithm is based on Euler's formula, as dedicated in Eq. (2). Phase subtraction is realized by negation and multiplication. We remove the gain in amplitude by dividing its average.

$$A_{ref}e^{-j\varphi_{ref}} = A_{ref}\cos\varphi_{ref} - jA_{ref}\sin\varphi_{ref} = I_{ref} - jQ_{ref} \quad (2)$$

$$A_{mea}e^{j\varphi_{mea}} \times A_{ref}e^{-j\varphi_{ref}} \times \frac{1}{A_{ref}} = \frac{A_{ref}}{A_{ref}}A_{mea}e^{j(\varphi_{mea}-\varphi_{ref})} = A_{mea}e^{j(\varphi_{mea}-\varphi_{ref})}$$

where $\left(\dfrac{A_{ref}}{A_{ref}} \approx 1\right)$

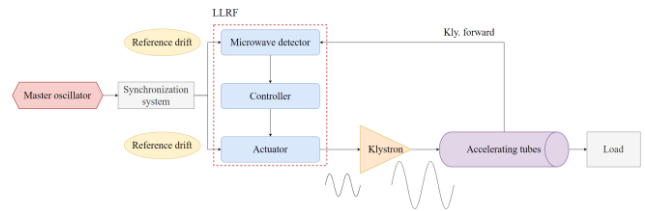

Figure 1: The structure of the microwave system in DCLS.

## FIRMWARE IMPLEMENTATION

Figure 2 illustrates the firmware implementation of the algorithm. There are eight radiofrequency detection ports and four ADCs. Detected microwave signals (including klystron forward, reference, and vector modulator) are down-converted and sampled by ADC (Analog-to-Digital Converter). BUFG generates the global clock (105 MHz) after clock recovery. The data recovery, demodulation, and filter are processed in FPGA. FIR filter2 and CIC filter have a narrower bandwidth to obtain the average reference amplitude. We also designed a switch to turn the algorithm function on or off. Finally, Vector I and Q are demodulated and output.

## EXPERIMENTS AND RESULTS

We conducted experiments and algorithm validation by combining the DCLS synchronization and LLRF system, shown in Figs. 3 and 4. The hardware of the DCLS LLRF is based on the MTCA. 4 standard. RF AFE (Analog Front-

___________________
* Work supported by the National Natural Science Foundation of China (Grant No. 22288201), the Scientific Instrument Developing Project of the Chinese Academy of Sciences (Grant No. GJJSTD20220001), and the Shenzhen Science and Technology Program (Grant No. RCBS20221008093247072).
† zhujinfu@mail.iasf.ac.cn
‡ dinghongli@dicp.ac.cn

End) is an independent chassis that processes reference signals from the synchronization system to generate LO (Local Oscillator), ADC sampling clock, and reference signals. Three signals are connected to the RTM board (DWC8VM1) for down-conversion. Digital sampling is realized on the AMC board (SIS8300). Vector modulator outputs excitation to drive SSA (Solid-State Amplifier).

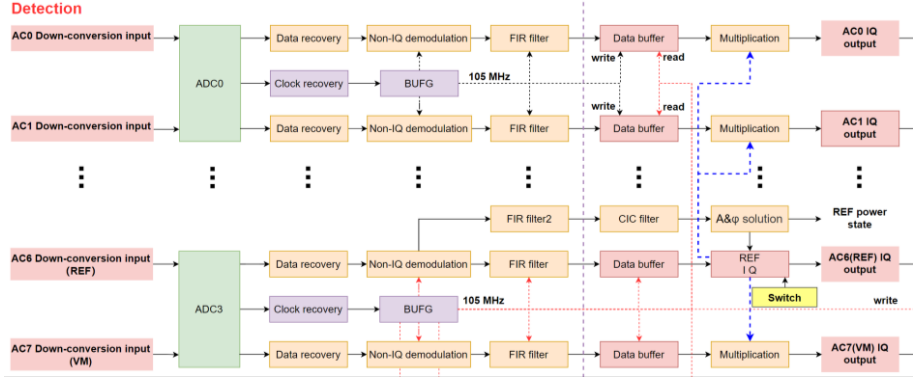

Figure 2: The model diagram of the DCLS device.

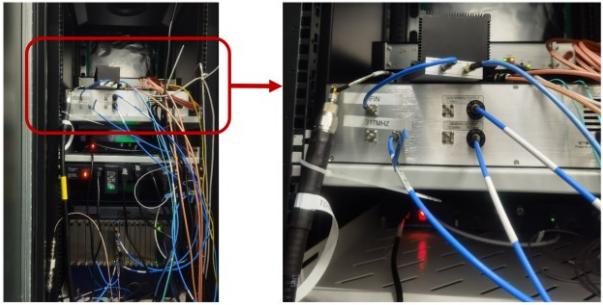

Figure 3: Actual object of experiments.

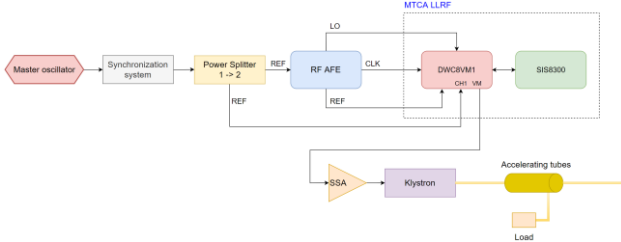

Figure 4: The block diagram of experiments

After the power split, another synchronization signal is connected to CH1 for detection. The experiment tested the amplitude and phase distribution within one hour when the algorithm function was turned off and on. As shown in Figs. 5 and 6, the phase stability (RMS) improved from 0.058 to 0.017 degrees, but the amplitude stability (RMS) declined from 0.035% to 0.063%, resulting from the amplitude multiplication in the algorithm. It causes irrelevant amplitude noise to overlap. Consequently, this reference tracking algorithm is suitable for control algorithms with low-delay requirements of microwave measurement.

## CONCLUSION

This paper introduces a low-delay reference tracking algorithm for microwave measurement and control. Experiments show that it can eliminate the drift in the LLRF clock allocation network, which causes the measurement drift in microwave detection. However, it also increases the amplitude measurement noise. Nevertheless, this reference tracking algorithm is suitable for control algorithms with low-delay requirements of microwave measurement.

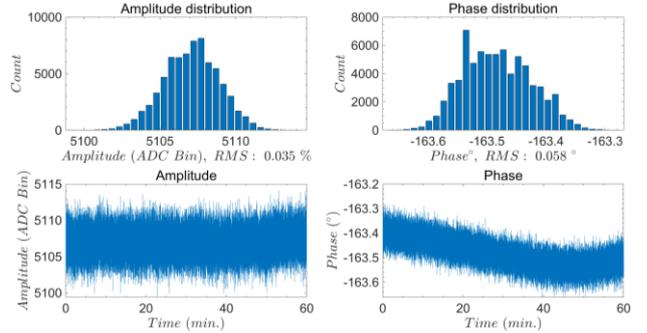

Figure 5: The Ch1 amplitude and phase without the algorithm.

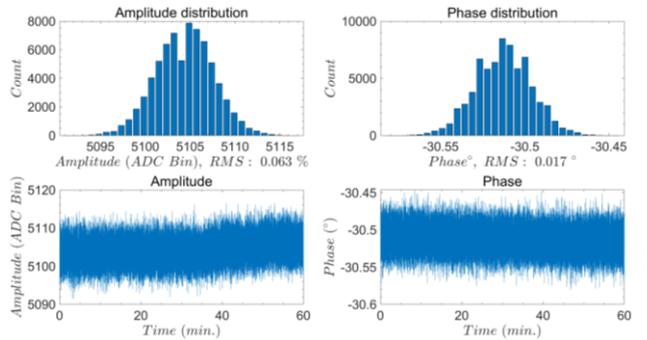

Figure 6: The Ch1 amplitude and phase with the algorithm.

## ACKNOWLEDGEMENTS

The authors would like to thank the Institute of High Energy Physics, Shanghai Advanced Research Institute, DESY, etc. who collaborated with the DCLS for their support and various discussions over the years.